\newcommand{\cm}{{~\rm cm}}
\newcommand{\s}{{~\rm s}}
\newcommand{\km}{{~\rm km}}

\newcommand{\erg}{{~\rm erg}}
\newcommand{\yr}{{~\rm yr}}

\newcommand{\kpc}{{~\rm kpc}}
\newcommand{\keV}{{~\rm keV}}

\documentclass[12pt,preprint]{aastex}
\shortauthors{Sternberg et al.}

\begin{document}

\title{INTERPRETING A DWARF NOVA ERUPTION AS MAGNETIC FLARE ACTIVITY}

\author{Noam Soker\altaffilmark{1} and Saeqa Dil Vrtilek\altaffilmark{2}}

\altaffiltext{1}{Department of Physics, Technion$-$Israel Institute of
Technology,
Haifa 32000, Israel; soker@physics.technion.ac.il}
\altaffiltext{2}{Harvard-Smithsonian Center for Astrophysics,
Cambridge, MA 02138; saku@head.cfa.harvard.edu}

\begin{abstract}
We suggest that the radio emission from the dwarf nova SS Cyg
during outburst comes from magnetic activity that formed a corona
(similar to coronae found in magnetically active stars), rather
than from jets. We base our claim on the recent results of Laor \&
Behar, who found that when the ratio between radio and X-ray flux
of accretion disks in radio-quiet quasars is as in active stars,
$L_r/L_x \la 10^{-5}$, then most of the radio emission comes from
coronae. Using observations from the literature we find that for
SS Cyg during outburst $L_r/L_x < 10^{-5}$. This does not mean
jets are not launched during outbursts.  On the contrary, if the
magnetic activity in erupting accreting disks is similar to that
in stars, then mass ejection, e.g., as in coronal mass ejection,
is expected. Hence magnetic flares similar to those in active
stars might be the main mechanism for launching jets in a variety
of systems, from young stellar objects to massive black holes.
\end{abstract}


\section{INTRODUCTION}
\label{sec:intro}

In a recent detailed and elegant study, Laor \& Behar (2008) present a
correlation
between radio and X-ray emission over some 15 orders of magnitude, from
magnetically active stars to radio quiet AGN.
The correlation they observe is consistent with $L_r = 10^{-5} L_X$, a
correlation
known as the G\"udel-Benz
relation in magnetically active stars (G\"udel \& Benz 1993). This suggests
that the source of the
radio emission in these highly diverse objects may be related.
Indeed Laor \& Behar (2008) suggest that in
radio quiet AGN the source of radio emission is coronae above accretion
disks.
Suzaku observations of the dwarf nova (DN) SS Cyg (Ishida et al. 2008)
provide spectral evidence for thermal plasma distributed on the disk during
outburst analogous to the Solar corona.
Suggestions for the presence of coronae above accretion disks are not new (e.g., Galeev et al.
1979; Done \& Osborne 1997; Wheatley \& Mauche 2005),
and the connection between coronae and jets were proposed in the past
(e.g., Fender et al. 1999; Markoff et al. 2005).
The results of Laor \& Behar (2008) and Ishida et al. (2008)
put the presence of coronae in accretion disks on solid ground.
The correlation does not hold for radio loud AGN, where
most of the radio emission comes from jets, and
systems show $L_r \gg 10^{-5} L_X$.
Laor \& Behar (2008) find that some galactic black hole (GBH) binaries also
reside close to this correlation.
This implies that the common practice of attributing radio emission
of accreting neutron stars and black holes (BH) to emission from jets
(e.g., Dunn et al. 2008) should be done with caution,
if we assume that these systems are analogous with AGN (e.g., Markoff
2006).
It is correct to attribute the radio emission to jets in some cases,
as in SS~433 (Miller-Jones et al. 2008), but not necessarily in all
binaries.

Jets are not usually observed in cataclysmic variables (CVs),
among them DN.
In CVs a white dwarf (WD) accretes mass from a companion via an accretion
disk.
In many other astrophysical systems accretion disks
are known to launch jets.
The absence of jets in CVs impose strong constraints on some jet
launching models (Soker 2007).
For example, no jets are observed in intermediate polars (DQ Her systems).
These are cataclysmic variables where the magnetic field of the accreting WD
is thought to truncate the accretion disk in its inner boundary.
This magnetic field geometry is the basis for some jet-launching
models in YSOs (e.g., Shu et al. 1991). Why then are no jets observed from
intermediate polars?

Theoretical arguments (Soker \& Lasota 2004), and at least one observation
(Retter 2004) claimed that jets might be present in CVs when the accretion
rate is high.
High accretion rates result in longer diffusion time for photons
(radiation) from the disk,
and might leave time for the energy to be channeled to kinetic
(Soker \& Lasota 2004) or magnetic (Soker 2007) energy.
Soker (2007) proposed that in some cases jets are launched in a manner similar
to coronal mass ejection (CME) in the sun.
Instead of a dynamo and buoyant magnetic flux tubes as in the sun, in
accretion
disks the kinetic energy of the gas in the accretion disk is transferred,
e.g.,
by shock waves, to thermal energy.
The thermal energy builds pressure that inflates magnetic flux loops
above the disk,
and from there on the activity is analogous to solar flares.
It is also possible that the
kinetic energy is transfered directly to the magnetic field.
This model predicts that some burst activity must precede the
magnetic activity.
The idea that jets are launched by magnetic fields reconnection events similar
to that in the sun is not new, e.g., de Gouveia Dal Pino \& Lazarian (2005;
de Gouveia Dal Pino 2006).

In a recent paper K\"ording et al. (2008) attributed radio emission
in an outburst of SS Cyg to jets.
Here we argue in $\S 2$ that the radio emission in SS Cyg is more likely to
come from a corona that was formed during the outburst.
The same activity could have launched jets, or more generally,
collimated outflows, as we argue in $\S 3$.
Fender et al. (1999; also Markoff et al. 2005) proposed that the base of the jet
and the corona are the same region. Our idea goes beyond this identification, and we argue
that the acceleration mechanism of the jet is similar to solar magnetic activity.
Our results are not in contradiction with the results of these works. Our claims are
aiming at identifying the acceleration process of the jet.
To put our claim on a broader view, in $\S 4$ we discuss some related
systems.

\section{THE OUTBURST BEHAVIOR IN SS CYG}
\label{sec:burst}
\subsection{Radio and X-ray fluxes}

K\"ording et al. (2008) conducted radio observation of SS Cyg in its
April 2007 outburst. The peak radio luminosity at 8.6~GHz reached a value
of $\sim 1~$mJy,
and then declined to $\sim 0.3$mJy in a short time.
It then continued slowly to decline. We take the average outburst
radio emission to be $\sim 0.3~$mJy.
This gives a radio flux of
$\bar F_{\rm radio} \simeq 3 \times 10^{-17} \erg \cm^{-2} \s^{-1}$.

No X-ray observations were conducted during the April 2007 outburst,
and K\"ording et al. (2008) refer to X-ray observations at previous
outbursts.
Wheatley et al. (2003) followed the X-ray emission from SS Cyg in its
October 1996
outburst and analyze the X-ray emission as coming from two sources.
Strong emission from the boundary layer (the boundary of the accretion disk and the WD),
with a maximum flux of
$L_{\rm BL} \simeq 3.6\times 10^{-10} \erg \cm^{-2} \s^{-1}$.
(The fluxes here include their correction of a factor of 1.8 to
include the total X-ray emission).
This emission is obscured, because of high optical depth, during most
of the outburst (Patterson \& Raymond 1985a, b). Wheatley et al.
identify another source, the residual,
which they interpret as extended emission, possibly coronal.
The residual X-ray emission flux is
$F_{ex}\simeq 2.5 \times 10^{-11} \erg \cm^{-2} \s^{-1}$.

Okada et al. (2008) find the X-ray luminosity of the September 2000 outburst
to be $\sim 0.3$ that in quiescence. Taking the quiescent flux from
Wheatley et al. (2003), the September 2000 outburst X-ray luminosity is
$\simeq 3 \times 10^{-11} \erg \cm^{-2} \s^{-1}$.
The ASCA flux in the $0.8-10 \keV$ band reported by  Baskill et al. (2005)
was higher than the quiescent one: $4 \times 10^{-11} \erg \cm^{-2}
\s^{-1}$
in the outburst versus $2 \times 10^{-11} \erg \cm^{-2} \s^{-1}$ in
quiescence.
Similar X-ray outburst fluxes were observed in the past (e.g., Nousek et al.
1994).

Over all, we find the ratio of radio to residual X-ray emission (the X-ray
emission that
is not attributed to the boundary layer by Wheatley et al. 2003) during
outburst to be
\begin{equation}
\frac{F_{\rm radio}}{F_{x}} \simeq 10^{-6}.
\label{eq:tkh1}
\end{equation}

For a distance of 166 \kpc~to SS Cyg (Harrison et al. 1999) the residual
X-ray luminosity is
$L_{x} \simeq 10^{32} \erg \s^{-1}$. We can now place the eruptive SS Cyg on
the radio vs. X-ray luminosity seminal plot ($L_r$ vs. $L_x$ plane) of Laor \&
Behar (2008).
We do this in Figure 1.
The radio and residual X-ray luminosities of SS Cyg at outburst
put it where the magnetically active stars are.
\begin{figure}  %
\vskip +3.5 cm
\begin{tabular}{cc}
{\includegraphics[scale=0.70]{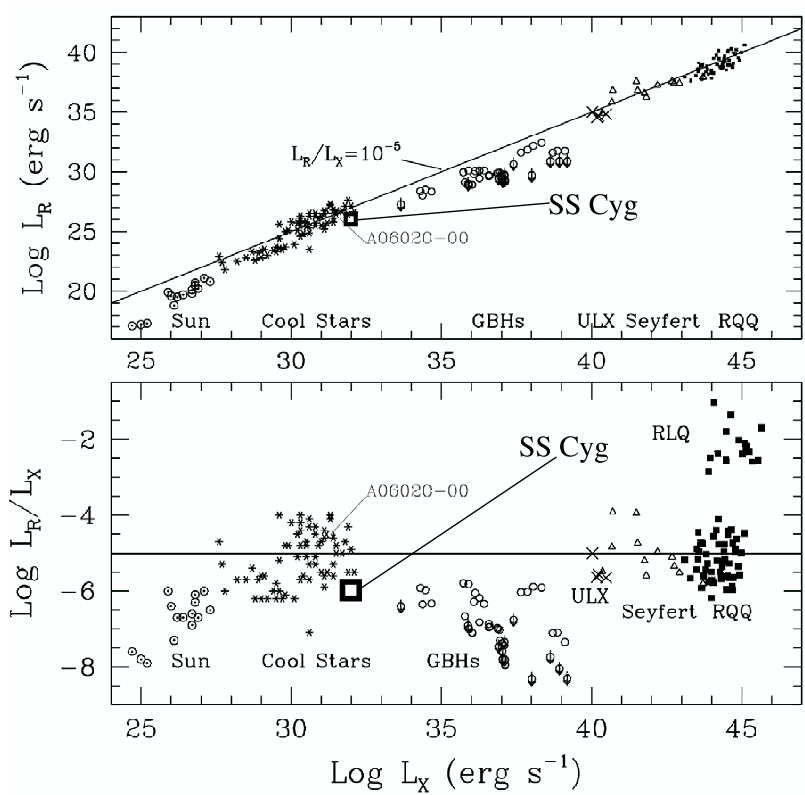}}
\end{tabular}
\vskip -4.5 cm
\caption{The position of different objects in the $L_r$ vs. $L_x$ plane
(from Laor \& Behar 2008, beside SS Cyg).
GBHs, ULX, and RQQ stand for Galactic black holes (in the hard state),
ultraluminous X-ray sources, and radio quiet quasars, respectively.
The square is the location of SS Cyg during eruption.
The location of SS Cyg in this plane hints that its eruption is connected to
magnetic flares.
}
\label{fig:models}
\end{figure}

The total flux at maximum, dominated by the extreme UV (Wheatley et al.
2003), is $F_{\rm tot} \simeq 3 \times 10^{-8} \erg \cm^{-2} \s^{-1} \sim 10^{3} F_{x}$.
A bolometric to X-ray ratio of $\sim 10^{3}$ is also observed in
magnetically very active stars, i.e., in the activity saturation regime (Pizzolato et al.
2003).
This further suggests that what we observe in the outburst of SS Cyg
is a magnetic outburst, similar to flares on very active stars.

\subsection{Time delay and kinetic energy}
Wheatley et al. (2003) find that for SS Cyg the X-ray rise occurs $\sim~1$day after the
visible. Most of the initial X-ray rise is due to the emission from the
boundary layer, and not from a corona. For that, we cannot tell when X-ray
emission
from the corona starts. The radio emission, which we attribute to the
corona, is delayed also by $\sim
1~$day after the visible  (K\"ording et al. 2008).
This suggests that coronal emission starts after an instability in the
disk has set in. In our model, kinetic and gravitational
energy released in the disk lead to magnetic activity, rather than
magnetic activity increasing the accretion rate.

In the solar case, the kinetic luminosity of the wind $\sim 2 \times 10^{27}
\erg \s^{-1}$
is about equal to the average X-ray luminosity.
Applying the same relation to SS Cyg, i.e., taking the kinetic outflow
to be equal to the residual X-ray luminosity, we get the kinetic energy of
the outflow to be $\sim 10^{-3}$ the energy released in the outburst.
If the outflow is equal to the Keplerian velocity at the WD surface,
$\sim 3000 \km \s^{-1}$, we find the mass loss rate to
accretion rate to be $\sim 10^{-3}$.
This ratio is typical for weak jets.
Namely, it is possible that weak jets are launched by the coronal magnetic
activity.

We further note that in some cases, like the BH candidate Cygnus X-1, the
power of the jets is
about equal to the bolometric X-ray luminosity (Russell et al. 2007).
This points to an ejection mechanism similar to that of
the solar wind, assuming that a large fraction (but not all)
of the X-ray emission is due to a corona.

The radio luminosity of Cyg X-1 is $10^{-8}$ times the X-ray luminosity,
which is typical for many GBH systems and some solar regions (Laor \& Behar
2008).
As this is the case for solar microflares, it is not unlikely that the
explanation
for the outflow is indeed a solar like activity.
The above ratio of $L_r/L_x \simeq 10^{-8}$ is much lower than in
magnetically active stars.
However, Wood et al. (2005) found that the kinetic energy of the wind
decreases
for more magnetically active stars.
The ratio of the average radio luminosity of the quiet sun (Drake et al.
1993)
to the average solar X-ray luminosity  is $\sim 3 \times 10^{-7}$.
Therefore, it is possible that the highest ratio of outflow kinetic energy
to X-ray luminosity
is obtained by activity similar to the less active regions of the suns,
where radio emission can gets as low as $L_r \simeq 10^{-8} L_x$ (G\"udel \&
Benz 1994).
Indeed, microflares can heat the corona and supply the energy to the wind
(Moore et al. 1999).

\section{THE AMPLIFICATION OF THE MAGNETIC FIELD IN THE OUTBURST}
\label{sec:dynamo}

Following the previous section and building on the behavior of the solar
magnetic field,
we consider magnetic fields with coherence lengths much smaller than the
radius of the
disk $r$, rather than the large scale magnetic field that is used in many
jet-launching models.
We take the magnetic pressure to be limited by the thermal pressure of the
disk,
$P_B \simeq P$, and approximate the thermal pressure from the hydrostatic
equation in the vertical
direction $z$: $d P/dz =-\rho (GM/r^2)(z/r)$, where $r$ is the radial
coordinate in
cylindrical coordinates, $\rho$ is the density, and $M$ is the mass of the
central accreting object.
This gives for the magnetic pressure
\begin{equation}
P_B \sim P \simeq \rho \frac{GM}{r} \left( \frac{H}{r} \right)^2,
\label{eq:pb}
\end{equation}
where $H = \epsilon r \sim 0.1r$ is the scale height of the disk in the $z$
direction.
The Alfven speed $v_A^2=(2P_B/\rho)$ is then
\begin{equation}
v_A \sim \frac{H}{r} v_{\rm esc} \ll v_{\rm esc},
\label{eq:va}
\end{equation}
where $ v_{\rm esc}= ({2GM}/{r})^{1/2}$ is the escape velocity from the disk
at radius $r$.
The inequality implies that the magnetic fields do not contain enough energy
to expel
large quantities of gas at high speeds.
If a corona is formed, the typical temperature associated with this
value of
$v_A$ is
\begin{equation}
T_{\rm corona} \sim 5 \times 10^6
\left( \frac{r}{R_{\rm WD}} \right)^{-1}
\left( \frac{H}{0.1r} \right)^{2} K
\label{eq:tcor}
\end{equation}
where for the WD mass and radius we took $M_{\rm WD}=1 M_\odot$ and $R_{\rm
WD}=0.01 r_\odot$,
respectively.
A hot corona can be marginally formed under these conditions, and only very
close to the
WD.

We turn now to an outburst, where the mass accretion rate substantially
increases.
Consider a rapid stochastic dissipation of the kinetic energy of the gas in
the disk.
This can occur through shock waves, i.e., transfer of kinetic energy to
thermal energy,
and through transfer of kinetic energy directly to magnetic fields.
In the case of shock waves the gas will cool via two processes.
Radiative cooling via diffusion of photons, and adiabatic cooling.
When the diffusion time scale is longer than the adiabatic expansion time
scale,
the orbital kinetic energy of the gas in the disk is transferred mainly to
vertical motion (perpendicular to the disk plane), that can lead to the
formation of jets
(Torbett 1984; Torbett \& Gilden 1992; Soker \& Regev 2003; Soker \& Lasota
2004).
This is the thermally launched jet model (Soker 2007).

Soker \& Lasota (2004) found the minimum accretion rate required to launch
jets from WDs in
the thermally launched jet model to be
\begin{equation}
\dot M_b \ga 10^{-6} \left( \frac {\alpha_d}{0.1} \right)^{-1} M_\odot
\yr^{-1},
\label{eq:mb}
\end{equation}
where $\alpha_d$ is the disk-viscosity parameter. Due to several
uncertainties,
the limit in equation (\ref{eq:mb}) can be as low as $\sim 10^{-7} M_\odot
\yr^{-1}$.
This limit is compatible with the result of Retter (2004), who argued for a
detection
of jets in the transition phase (few months post-outburst) of nova
V1494 Aql.
Considering that $\alpha_d$ can be much lower, and taking into account some
observations
of jets in super-soft X-ray sources (see discussion in Soker \& Lasota
2004),
this limit can be as low as $\sim 10^{-8} M_\odot \yr^{-1}$.
The accretion rate at peak luminosity in the eruptions of SS Cyg is
$\sim 10^{-8} M_\odot \yr^{-1}$ (Wheatley et al. 2003).
This limit implies that if rapid dissipation of the orbital kinetic energy
occur,
then a large fraction, or even most, of this energy will be channelled to
accelerate gas in the vertical direction.

The vertically accelerated gas can stretch magnetic field lines and amplify
the magnetic fields (Soker 2007).
The strong magnetic fields lead to a second
acceleration stage, where after the primary outflow stretches magnetic
field lines,
the field lines reconnect and accelerate small amount of mass to very high
speeds.
This double-stage acceleration process might form highly relativistic jets
from BH
and neutron stars, as well as jets from brown dwarfs and stars.

Here we raise the possibility that the magnetic fields might be amplified by
an
$\alpha \Omega$ dynamo, in addition to a pure stretching in the vertical
direction.
The amplification of magnetic fields by turbulence dynamo, and the acceleration of gas
by reconnection in accretion disk coronae was considered before by
de Gouveia Dal Pino \& Lazarian (2005).
Here we attribute a major role to the differential rotation velocity and the dissipation
of the kinetic energy of the disk material. We also do not require the presence of a global
field, but rather consider many small scale flares.

In stars, the $\alpha$ effect (not to confuse with the $\alpha_d$ for disk
viscosity)
is attributed to the convection, while in galaxies it can come from
supernovae and other stars,
or from the dynamics of magnetic field lines themselves (Beck et al. 1996).
Basically, local heating (e.g. OB associations and supernova) in the galaxy
can drives the
$\alpha$ effect.
Here the local heating is done by small shocks (Soker \& Regev 2003), which
can then grow
to larger volumes. This result in a rapid increase in the thermal pressure,
and the acceleration of gas, mainly in the vertical direction, but locally
in other directions.
This motion can play the role of the $\alpha$ effect.
Within the disk the accelerated gas move at a fraction of the Keplerian
speed (only the outer layers can be accelerated to the escape speed):
$v_t = \beta v_{\rm Kep}$.
If we take the coherence length of the field to be
$H=\epsilon r$, then $\alpha ={\rm min}(\Omega H, v_t)$ (Beck et al. 1996,
sec. 4.4).
If $\beta \ga \epsilon \simeq 0.1$, then the magnetic growth time is (Beck
et al. 1996, sec. 4.4)
\begin{equation}
\tau_B \simeq \left( \frac{H}{\alpha \Omega} \right)^{1/2} \simeq
\Omega^{-1}.
\label{eq:taub}
\end{equation}

In the eruption phase the Alfven speed is about equal to the Keplerian
speed, and by
equation (\ref{eq:tcor}) (taking $H \simeq r$) a very hot corona is formed
up to a distance of
$r \simeq {\rm few} \times 10 R_{\rm WD}$ (somewhat cooler corona can be
formed at
larger distance as well).
The magnetic field growth time at these radii is
$\tau_B  \simeq \Omega^{-1} \simeq 5 (r/30 R_{\rm Wd})^{3/2}$~min.
This is much shorter than the rise time of the outburst at different bands
(spanning several hours to a day; K\"ording et al. 2008; Wheatley et al.
2003).

To summarize, in this section we examined a chain of events starting with
some disk instabilities that lead to enhanced mass accretion rate (Lasota
2001).
Local dissipation of orbital kinetic energy of the disk material, e.g., via
shocks, lead to
local expansion of many regions in the disk (Soker \& Regev 2003), because
radiative
cooling proceeds on a time scale longer than the expansion time (Soker \&
Lasota 2004).
The gas motion in the disk can amplify magnetic fields by stretching
magnetic field lines,
or by being the source of the $\alpha$ effect in an $\alpha \Omega$ dynamo.
These magnetic fields are likely to behave similarly to stellar magnetic
fields (Laor \& Behar 2008). Namely, they can be the main radio source and can
eject material by reconnecting in the disk corona (de Gouveia Dal Pino \& Lazarian 2005).
When collimated, the ejected gas becomes jets.

\section{RELATED OBJECTS}
\label{sec:objects}

Several authors (e.g. Nipoti et al 2005; Massi 2005) have pointed out
that X-ray binaries, and in
particular the subset within them that are called microquasars,
show evidence for radio-loud and radio-quiet states similar to that
observed in AGN.
The obvious analogy is that radio-loud systems emit strong jets and
radio quiet systems do not.
SS 433 is a complicated object, but shows that in cases where
jets are strong radio sources the ratio of radio to X-ray emission
can be $L_r/L_x \gg 10^{-5}$.
On July 11, 2003, the ratio of radio to X-ray emission from the core was
$(L_r/L_x)_{\rm core} \simeq 2 \times 10^{-5}$
(Migliari et al. 2005; Miller-Jones et al. 2008).
The radio contribution outside the core was $0.54$ times that in the core.
At other times X-ray and radio measurements are not performed
simultaneously,
but we note that  the core X-ray emission is $\sim
0.025-0.25$
times that in July 11, 2003 (Migliari et al. 2005) ,
while the total radio emission is larger (Miller-Jones et al. 2008) by
factor of $\sim 2$.
Over all, for most of the time in SS 433, the radio to X-ray ratio is
$L_r/L_x \sim 10^{-4}-10^{-3}$.
This ratio resides between radio quiet and radio loud AGN.

In Table 1 we list the $F_r/F_x$ ratios of a few  X-ray binaries
associated with micro-quasars that are listed as ``radio-loud" by Massi (2005).
We determine the majority of our ratios from numbers obtained from Figures 2 and 6
of Gallo, Fender, \& Pooley (2003; hereafter GFP03).
Where available, for each object we give the range from each of the the states: quiescent, hard,
soft, and transient.  GFP03 note that the quiescent state is dominated by
jets, no jets are observed in the hard state, and jets again in the soft and transient state.
The majority of the ratios do lie below 1.e-5 as noted by Loar \& Behar (2008).
Nevertheless, as GFP03 note, many of these systems do show jets.
The reason is that the accretion disks around neutron stars and GBHs has
a substantial fraction of their emission in the X-ray band. Disks around
stars and massive BH emits mainly in the IR to UV bands.
The strong thermal X-ray emission from disks around neutron stars and GBHs
(X-ray binaries) reduces the ratio of $L_r/L_x$.
The inner region of disks around WDs contribute to the X-ray emission,
particularly from the boundary layer.
The `trick' in SS Cyg in eruption is that the optical depth to the boundary layer becomes
very large, such that the energy is radiated from a larger area at longer wavelengths.
This allows the detection of the residual X-ray emission, which is assumed to come
from the coronae.

Not only are the disks in X-ray binaries strong x-ray emitters, but the observed luminosity
is strongly influenced by the geometry of the system. This is because the flux comes from
a small region towards the center of the disk that can be easily obscured by flared
edges of the disk, if observed edge on.
We conclude that the observed X-ray flux from X-ray binaries is not a reliable measure
of the X-ray flux from the disk corona.
Hence the ratio $F_r/F_x$ can not be used as an indicator for jet formation.


\begin{table}

{\bf Table 1: Related Objects}

\footnotesize
\bigskip
\begin{tabular}{lcccc} \hline

Source& Fr/Fx& State&Jets&Reference\\
\hline
\hline
V404 Cyg&5.2e-5&quiescent&yes&Gallo et al. 2003\\
&4.0e-6-1.0e-5&hard&no&Gallo et al. 2003\\
&&&\\
GX339-4&1.2e-5-6.3e-6&quiescent&yes&Gallo et al. 2003\\
&2.1e-8-2.7e-6&hard&no&Gallo et al. 2003\\
&&&\\
XTEJ1118+480 &   1.5e-6 &outburst&   yes         &   Chaty et al. 2003\\
 & 2.4e-6&hard&no&Gallo et al. 2003\\
&&&\\
Cygnus X-1&7.5e-8-7.5e-7&hard&no&Gallo et al. 2003\\
&&&\\
Cygnus X-3&1.0e-7-1.5e-5&soft&yes&Gallo et al. 2003\\
&&&\\
GRS 1915+105&2.0e-9-5e-7&transient&yes&Gallo et al. 2003\\
&&&\\
SS433        & 1.0e-3-1.0e-4 & & yes-steady &  Miller-Jones et al 2008\\
            &   1.0e-4 &   &yes        &   Kotani et al. 1999\\
\hline
\hline
\end{tabular}
\footnotesize
\bigskip

\normalsize
\end{table}

\section{SUMMARY AND DISCUSSION}
\label{sec:summary}

Magnetically active stars follow the
G\"udel-Benz relation ($L_r \simeq 10^{-5} L_X$;
G\"udel \& Benz
1993).
In a recent paper Laor \& Behar (2008) found this correlation
to hold over some 15 orders of magnitude, from magnetically active stars to
radio quiet AGN.
For strong jets to be present in AGN $L_r \gg 10^{-5} L_X$.

The new finding of Laor \& Behar (2008), and the observations of Ishida et al.
(2008) of disk coronae formed during an outburst of SS Cyg,
motivated us to examine whether the radio emission found in the
outburst of the dwarf nova SS Cyg by K\"ording et al. (2008) is due
to magnetic flaring as well.

In accreting WD and neutron stars the task of finding the ratio $L_r/L_x$
is more complicated, as the accreting gas near the boundary layer emits in
the X-ray band.
As we showed in section \ref{sec:objects}, the strong X-ray emission from the
disks in X-ray binaries prevent us from using the ratio $L_r/L_x$ to learn about the
presence of absence of a jet.

However, in the case of SS Cyg
the high accretion rate during outburst obscures the boundary layer, and the
residual X-ray emission observed by Wheatley et al. (2003) can be attributed
to a corona.
In $\S 2$ we found the ratio between the peak radio emission at outburst
(K\"ording et al. 2008)
and the residual X-ray emission at outburst (Wheatley et al. 2003) of SS Cyg
to be
$L_r/L_x \simeq 10^{-6}$.
Note that the two measurements are in two different outbursts, but the
outbursts peaks
of SS Cyg are quite regular (Wheatley et al. 2003).

This finding leads us to suggest that most of the radio emission in the
outburst
of SS Cyg comes from a magnetically newly formed corona, and not from jets
as
argued by K\"ording et al. (2008).
This does not mean that jets, or a collimated outflow, were not launched in
the outburst.
On the contrary, if the magnetic activity on the surface of accretion disks
is similar to stellar magnetic activity, then ejection of mass is expected
(e.g., de Gouveia Dal Pino  \& Lazarian  2005).
Soker (2007) suggested that jets are launched in a manner analogous
to stellar magnetic eruptions and coronal mass ejection.
We have elaborated on this idea in $\S 3$.
We do note that Fender et al. (1999) and Markoff et al. (2005) proposed that
the base of the jet and the corona are the same region. We are not in dispute
with their claim. We simply argue that processes similar to solar flares form
the corona and accelerate the jets (de Gouveia Dal Pino \& Lazarian 2005),
and that this magnetic activity is the source
of the radio emission in SS Cyg, rather than the already collimated outflow (jets).

The main assumption of the stellar-like magnetic activity is that local
heating occurs
in the regions of the disk close to the accreting object, in a manner
described by
Soker \& Regev (2003):
if the accretion rate is high enough
such that
radiative cooling occurs on a time scale longer than the expansion time
scale,
then the gas motion due to the local heating is mainly in the vertical
direction.
The critical mass accretion rate for WDs is
$\sim 10^{-8}-10^{-7} M_\odot \yr^{-1}$ (Soker \& Lasota 2004).

The vertical motion can stretch magnetic field lines and amplify the
magnetic fields
(Soker 2007). Here we considered the $\alpha \Omega$ dynamo
(see also de Gouveia Dal Pino \& Lazarian 2005 who considered a turbulence dynamo).
Local velocities can be in all directions, not only vertical.
We assumed that such a motion can play the $\alpha$ role in the $\alpha
\Omega$
dynamo mechanism, and concluded that the time scale for the amplification of
the
magnetic field is very short compared with the rise time of the outburst.
It is very likely that the strong magnetic fields behave similarly to
stellar
magnetic fields (Laor \& Behar 2008).
We argued that these magnetic fields emit most of the radio emission observed
by K\"ording et al. (2008) in the eruption of SS Cyg.
The magnetic flares can also lunch jets, and might even be the main
mechanism for
lunching jets in variety of objects, from YSOs to massive BHs (Soker 2007).

\acknowledgements

This research was supported by the Asher Fund for Space
Research at the Technion, the Israel Science foundation, and a Smithsonian
short term visitor grant to Soker and NSF grant AST--0507637 and NASA grant
NNX08AJ61G to SDV.

\end{document}